\documentclass[twocolumn,english]{IEEEtran}
\usepackage{color}
\usepackage{babel}
\usepackage{array}
\usepackage{float}
\usepackage{graphicx}
\usepackage[unicode=true,
 bookmarks=true,bookmarksnumbered=true,bookmarksopen=true,bookmarksopenlevel=1,
 breaklinks=false,pdfborder={0 0 0},backref=false,colorlinks=false]
 {hyperref}
\hypersetup{pdftitle={Your Title},
 pdfauthor={Your Name},
 pdfpagelayout=OneColumn, pdfnewwindow=true, pdfstartview=XYZ, plainpages=false}

\makeatletter

\providecommand{\tabularnewline}{\\}

\ifCLASSOPTIONcompsoc
\usepackage[caption=false,font=normalsize,labelfont=sf,textfont=sf]{subfig}
\else
\usepackage[caption=false,font=footnotesize]{subfig}
\fi
\usepackage{textcomp}
\PassOptionsToPackage{warn}{textcomp}
\usepackage{url}

\makeatother

\begin{document}

\title{On the likelihood of multiple bit upsets in logic circuits}

\author{\authorblockN{Nanditha P. Rao, Shahbaz Sarik and }\authorblockN{Madhav
P. Desai}\\ \authorblockA{ Indian Institute of Technology \textendash{}
Bombay, Powai, Mumbai \textendash{} 400076, India\\
Email: \{nanditha@ee, shahbaz@ee, madhav@ee\}.iitb.ac.in}}
\maketitle
\begin{abstract}
Soft errors have a significant impact on the circuit reliability at
nanoscale technologies. At the architectural level, soft errors are
commonly modeled by a probabilistic bit-flip model. In developing
such abstract fault models, an important issue to consider is the
likelihood of multiple bit errors caused by particle strikes. This
likelihood has been studied to a great extent in memories, but has
not been understood to the same extent in logic circuits. In this
paper, we attempt to quantify the likelihood that a single transient
event can cause multiple bit errors in logic circuits consisting of
combinational gates and flip-flops. In particular, we calculate the
conditional probability of multiple bit-flips given that a single
bit flips as a result of the transient. To calculate this conditional
probability, we use a Monte Carlo technique in which samples are generated
using detailed post-layout circuit simulations. Our experiments on
the ISCAS'85 benchmarks and a few other circuits indicate that, this
conditional probability is quite significant and can be as high as
0.31. Thus we conclude that multiple bit-flips must necessarily be
considered in order to obtain a realistic architectural fault model
for soft errors.\end{abstract}

\section{Introduction}

\IEEEPARstart{ R}{eliability} of semiconductor devices has become
an important area of concern, especially with technology scaling.
Circuits are subject to permanent faults, caused due to transistors
failing because of phenomena such as hot carrier injection, negative
bias temperature stability, oxide breakdown and so on \cite{reliability_issues_scaling}.
Circuits can also be subject to transient faults caused due to process
variations, signal integrity issues, power supply noise and high-energy
particle strikes (neutrons from cosmic rays, alpha particles) \cite{reliability_permanent_transient}.
Such errors which do not cause permanent damage to the device are
called soft errors and the rate at which they occur is known as the
soft error rate (SER).

Soft errors have become a major reliability concern for semiconductor
devices in the past few decades. When high energy particles such as
alpha particles, protons or neutrons from either cosmic rays or packaging
materials strike sensitive regions of a semiconductor device, they
generate electron-hole pairs. This process leads to deposition of
a certain amount of charge in the device. For example, if the high
energy particle strikes a memory cell and if the deposited charge
is greater than a certain amount of charge called the critical charge,
it can flip the stored bit ($0\rightarrow1\, or\,1\rightarrow0$).
Since this error can be reversed by rewriting the correct value and
since it is not caused due to permanent physical damage to the device,
it is termed ``soft error''. A similar process can occur when a
particle strikes a sensitive p-n junction in a transistor, creating
a transient or a glitch in a combinational logic circuit.

A complete historical review and the experimental results of soft
errors and their impact have been presented in \cite{SEU_expts_IBM,SEU_expts_satellite}.
Experiments and field tests conducted on large computer systems at
different locations indicate a memory error rate of $1.6\times10^{-12}\,$upsets/bit-hr
and an average of one error per month \cite{gnd_level}. Most early
studies considered bit-flips in memories. \textcolor{black}{Several
studies have also indicated that the SER of a system tends to increase
as technology scales \cite{SER_trends1,SER_trends3_combn,masking_scaling_trends_phy},
which is mainly attributed to the increase in memory density.} Also,
as the technology scales, a single particle strike can affect more
than one memory cell at a time, resulting in multiple bit errors \cite{mult_bit_mem2,mult_bit_mem1}.

Single event transients (SETs) in logic circuits were not given as
much importance, because such circuits tend to exhibit inherent masking
phenomena (described in Section \ref{sec:Related-Work}), which prevent
the transients from causing an error at a latching element. Soft errors
are generally modeled by a single bit-flip fault model, in which,
a single random bit is expected to flip \cite{single_bit1,single_bit_flip_element,single_flip_arch}.
But, as the technology scales, a single particle strike can affect
more than one drain of a transistor, resulting in multiple transients
which can propagate to cause multiple bit-flips at the output of a
circuit \cite{Mult_glitch_sing_strike2}. Further, the masking phenomena
seem to reduce as technology scales, and thus, the impact of soft
errors on logic circuits is likely to increase, as has been indicated
in \cite{combn1_glitch_ht,MBU_combn,combn_seq,combn3_algo}.

In this paper, we address the following question: Suppose that a particle
strike affects a drain of a transistor resulting in a single transient
in a logic circuit. Given that this transient propagates and flips
the content of a single flip-flop, what is the probability that it
can flip multiple flip-flop values? This involves the computation
of a conditional probability of multiple bit-flips given that at least
one bit has flipped. If this conditional probability is significant,
then an architectural level fault model for soft errors must incorporate
multiple bit failures in order to identify effective error correction
schemes. A few studies done in the past, to quantify the multiple
bit-flip probability \cite{correlations1,correlations2,multiple_errors_archlevel}
indicate that multiple bit flips are likely. However, the studies
use approximate modeling techniques to arrive at these conclusions,
and the confidence in their conclusions is doubtful. Our work uses
a Monte Carlo scheme in which detailed simulations (using SPICE) of
post layout circuit netlists are used to arrive at estimates of this
conditional probability. Our results (on the \textit{ISCAS'85} benchmarks
and a few other circuits) indicate that, the probability of multiple
bit-flips amongst the faulty cases is significant and can be as high
as $0.31$. Thus, multiple bit-flips must be incorporated in order
to obtain a realistic fault model for soft errors.

The rest of this paper is organized as follows. In Section \ref{sec:Related-Work},
we review related work on fault models for soft errors and prior work
on multiple bit upsets. Section \ref{sec:Methodology} describes our
methodology for calculating the probability of multiple bit flips.
Experimental results on \textit{ISCAS'85} benchmarks and a few other
circuits are presented in Section \ref{sec:Results}. Section \ref{sec:Conclusions}
summarizes and concludes the paper.

\section{\label{sec:Related-Work}Related Work}

Several experiments on memories have already demonstrated that multiple
bit upsets do occur due to a single particle strike \cite{mult_bit_mem2,mult_bit_mem1}.
Experiments in \cite{mult_bit_mem2} conclude that about 7\% of the
total events are multiple bit upsets (MBU) in memories. Schemes such
as bit interleaving along with error correction codes are already
in place to handle MBUs in memories. The rate of MBUs is reported
to increase with technology scaling, both in memories \cite{mult_bit_mem2}
and in logic circuits \cite{mult_combn_scaling}.

The impact of soft errors on combinational/sequential circuits was
not studied in detail in the past, due to the presence of the following
three inherent masking phenomena in circuits: 
\begin{enumerate}
\item Logical masking, in which, the type of logic gate and its input combinations
could mask the propagation of the transient. 
\item Electrical masking, in which the glitch can get attenuated as it passes
through the logic stages. 
\item Latching window masking, in which, the glitch might be masked if it
does not fall within the setup/hold times of the capturing latch. 
\end{enumerate}
However, as the technology scales, the impact of these masking phenomena
seems to reduce and hence the SER in logic circuits tends to increase
\cite{combn1_glitch_ht,MBU_combn,combn_seq,combn3_algo}. Several
algorithms have been designed to estimate SER of logic circuits in
general, without emphasizing on bit correlations or multiple bit upsets.
Some of these methodologies build models for glitch generation/propagation
and masking \cite{masking_systemFIT_algo,mars_SER_algo_hardng,SEAT_LA_SER_estimation},
employ testing and probability theory \cite{seq1_algo,observ_algo},
employ the method of binary decision diagrams \cite{BDD_SER_Miskov-Zivanov}
and probability transfer matrices \cite{prob_matrix_ADD_SER}, instead
of running circuit simulations. Most of these techniques are designed
to give fast estimates of the SER at the expense of accuracy. For
example, in \cite{static_algo_SER_estimation}, the authors report
an inaccuracy of up to 20\% against SPICE simulations. In \cite{masking_systemFIT_algo},
an SER estimation tool is built, under the assumption that logic paths
do not attenuate the glitches to a large extent and report an error
of 25\% in the results based on their assumption. They conclude that
single bit-flips account for over 95\% of all the bit-upsets in combinational
logic.

There have also been studies that aim to find the probabilities of
multiple errors caused due to a single particle strike. These studies
are mostly based on the fact that, as technology scales, a single-particle
strike can affect more than one drain of a transistor, resulting in
multiple transients, which can then propagate to cause multiple bit
upsets at the output of a circuit \cite{Mult_glitch_sing_strike2,mult_bit_mem2,mult_bit_mem1}.
There are also studies \cite{correlations1,correlations2} that estimate
the multiple bit-flip probability due to a single transient in the
circuit and show strong bit correlations. However, their methodology
is based on algorithms and approximate models for fault propagation,
and the corresponding inaccuracies reduce the confidence in their
conclusions. In \cite{multiple_errors_archlevel}, a fault-injection
experiment is conducted on a Verilog model of an embedded microprocessor
and they conclude that multiple bit errors occur in more than 30\%
of the faulty cases. However, their Verilog model does not take into
account the phenomenon of electrical masking. This can significantly
vary the results of the estimated SER and the rate of multiple bit-flips.
In \cite{fanout_multiple_paths2}, it is assumed that, a gate can
possibly affect multiple outputs in its fanout cone, but this assumption
is neither verified nor quantified. In \cite{correlations3_mult_upsets},
a transient fault simulator based on pulse and delay models is built.
Their results indicate that multiple bit-flips range from 0.9\% to
8.8\% of the injected faults and conclude that a single-bit flip model
``may not'' be adequate. However, they fail to consider the phenomenon
of electrical masking and also do not inject faults at the internal
nodes of logic gates. In \cite{multibit_gatemodel}, the authors conclude
that nearly 17\% of faults in combinational logic lead to multiple
bit errors. However, their analysis is based on gate level models
of the designs with timing delays, which is again an approximate methodology.

In order to increase the confidence in any conclusions about the likelihood
of multiple bit flips in logic circuits, it is thus essential that
the estimation technique be as accurate as possible. In order to improve
the accuracy, we use a Monte Carlo scheme combined with post-layout
circuit simulations in SPICE. Since we perform SPICE simulations without
making any simplifying assumptions (such as off-path logic values,
delay models etc.), all the masking phenomena in logic circuits are
taken into account, thereby increasing the confidence in the conclusions
of our experiments. There are performance issues, but our results
indicate that, with the use of parallel processing, non-trivial circuits
can be analyzed in a reasonable amount of time.

\section{\label{sec:Methodology}Methodology}

\begin{figure}[H]
\includegraphics[scale=0.49]{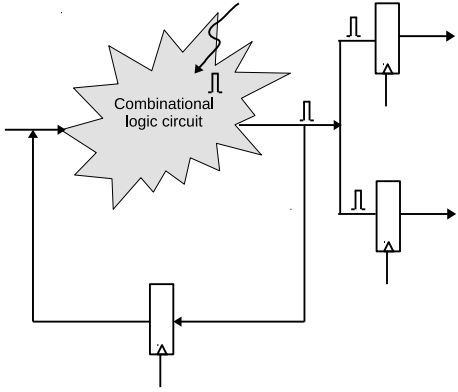}\caption{\label{fig:Experimental-setup}Experimental setup}
\end{figure}

The overall idea of our experiment is depicted in Fig. \ref{fig:Experimental-setup}.
Suppose that a single particle strike in a logic circuit results in
a single glitch or a transient. This can potentially propagate to
multiple outputs and cause multiple bits to flip, if it overcomes
the three masking phenomena described in Section \ref{sec:Related-Work}.
We intend to calculate the following conditional probability of multiple
bit-flips:

$P(Multiple\: bits\: flip\mid Single\: bit\: flips)$ as a result
of a single transient caused by a particle strike.

\begin{figure}[H]
\includegraphics[clip,scale=0.5]{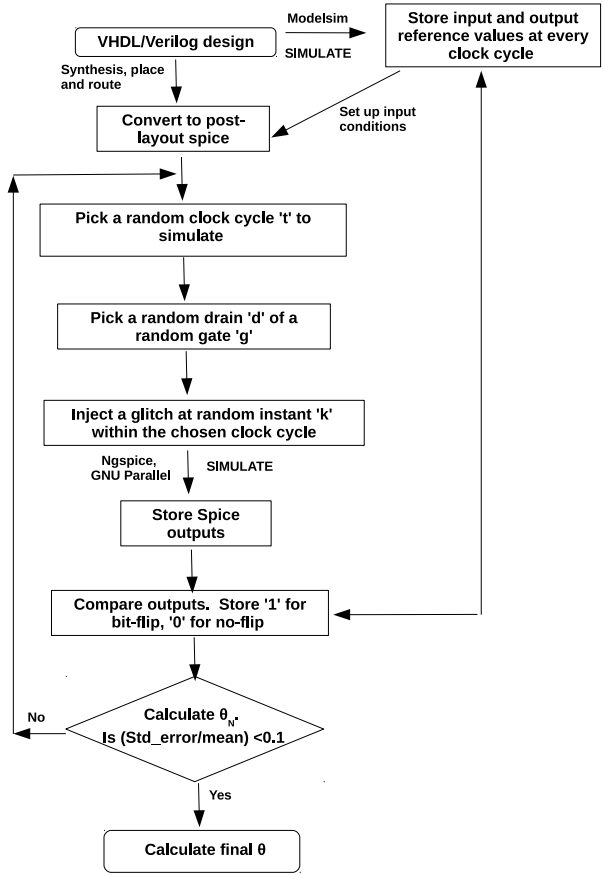}

\caption{\label{fig:Overall-flow-of}Experimental Methodology}
\end{figure}

Suppose that the total number of cases in which atleast one bit flips
is $N$ and of these, the number of cases in which multiple bits have
flipped is $N_m$, then we can estimate the conditional probability,
say, $\theta$ by the estimator 
\begin{equation}
\theta_{N}\ =\ \frac{N_{m}}{N}\label{eq:1}
\end{equation}
 The standard error in this estimate can itself be approximated as
\begin{equation}
\sigma_{N}\ =\ \sqrt{\frac{\theta_{N}-\theta_{N}^{2}}{N-1}}\label{eq:standardError}
\end{equation}

so that, with 95\% confidence, the actual value of $\theta$ lies
in the interval $[\theta_{N}-2\sigma_{N},\theta_{N}+2\sigma_{N}]$.
In our experiments, we continue generating samples until the standard
error is reduced to less than 10\% of the value of the estimate. The
quantity $\theta$ is calculated for several test circuits. The overall
flow of our methodology is depicted in Fig. \ref{fig:Overall-flow-of}.

A reference base is generated by first running an RTL simulation of
the test circuit using a representative test-bench. The test circuit
is then implemented to layout using synthesis tools, and the post-layout
circuit netlist is extracted (with parasitic capacitances). Sample
circuit simulation netlists are then generated from the post layout
netlist with a glitch injected at a random gate, and at a random point
in time. The inputs to the sample circuit simulation netlist are generated
by using the corresponding values from the reference base simulation
for the clock cycles corresponding to the sample time. The results
of the sample circuit simulation are compared with the reference outputs
from the reference base RTL simulation to check for bit-flips. Results
of several such sample simulations are tabulated to calculate an estimate
for $\theta$.

\subsection{Reference base generation}

As shown in Fig. \ref{fig:Overall-flow-of}, we start with a test
circuit which is described in RTL (VHDL or Verilog) together with
a reference test-bench. This RTL description is then implemented to
layout using synthesis tool (\textit{Synopsys Design Compiler} \cite{synopsys})
and place-and-route tool\textit{ (Cadence SoC Encounter} \cite{cadence}).
The post layout Verilog netlist is then simulated (with the reference
test-bench) using the \textit{ModelSim} \cite{modelsim} simulator
and the circuit inputs, outputs and flip-flop contents are recorded
at every clock cycle. These will be used as a reference base for generating
the sample circuit simulations, as well as for gathering bit-flip
statistics.

\subsection{Glitch injection}

When a high energy particle strikes a silicon substrate, the amount
of charge that it deposits, depends on the energy of the particle
and the node capacitance. The energy of the particle can vary between
10MeV to 100MeV \cite{combn1_glitch_ht}. The deposited charge determines
the magnitude of the transient current generated at the node. A particle
strike is most widely represented as a current source that is injected
at the drain of a transistor \cite{double_exp_drain_cur1,drain_cur2}.
Experiments conducted in \cite{combn1_glitch_ht} indicate that, for
a 130nm PMOS transistor, a particle of 10MeV energy can generate a
transient current of about 1.8mA. The transient current is mostly
modeled as a double exponential pulse \cite{double_exp_drain_cur1}:
\begin{equation}
I(t)\:=\:\frac{Q}{\tau_{a}-\tau_{b}}(e^{-t/\tau_{\alpha}}-e^{-t/\tau_{\beta}})
\end{equation}

where, Q is the deposited charge, $\tau_{\alpha}$ and $\tau_{\beta}$
are time constants that depend on process parameters.

\begin{figure}[H]
\includegraphics[scale=0.55]{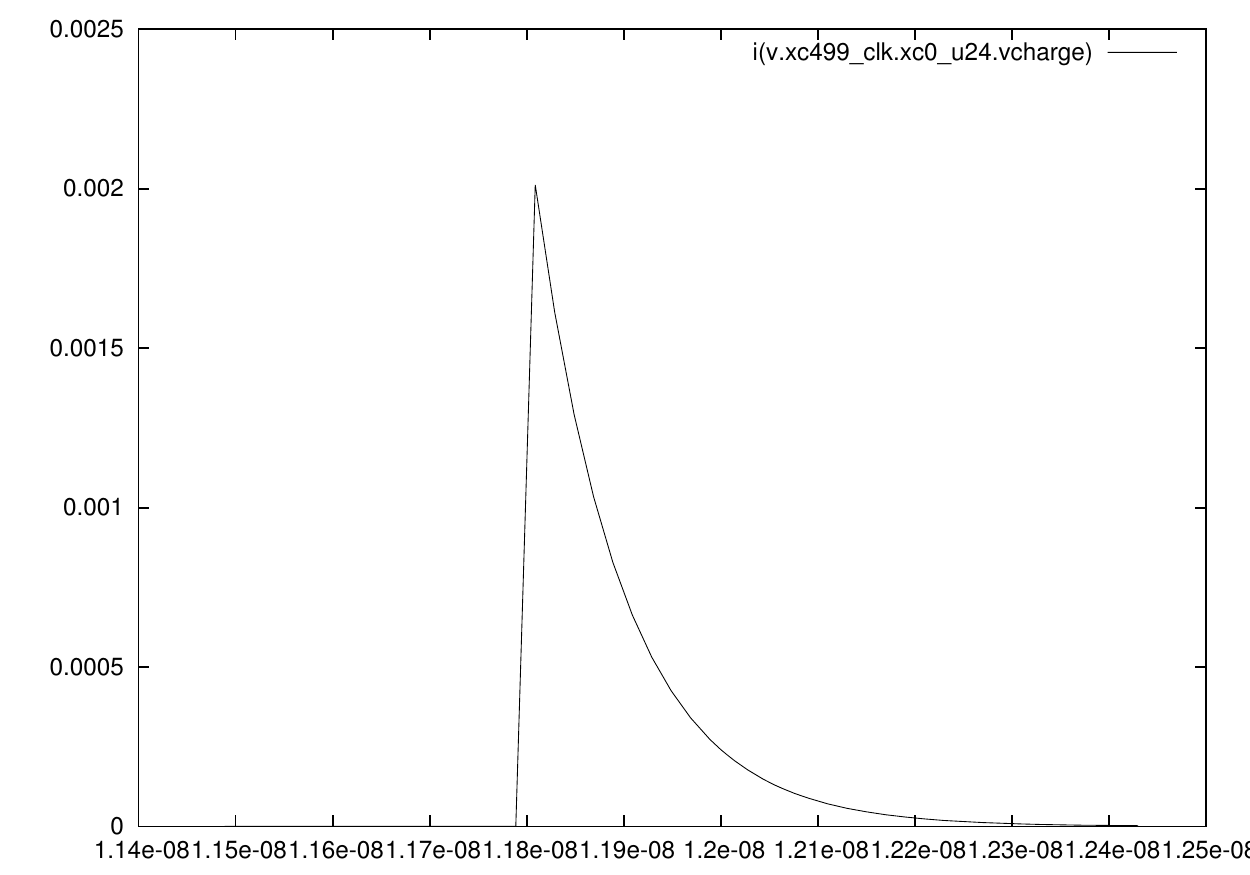}

\caption{\label{fig:Glitch-model}Glitch model}
\end{figure}

Based on the data reported in \cite{masking_scaling_trends_phy,combn1_glitch_ht},
scaled to our 180nm technology, we use the double exponential glitch
model shown in Fig. \ref{fig:Glitch-model}. A fixed glitch is used
in our experiments, though in reality, the glitch magnitude and decay
time depend on the particle energy, location of the particle strike,
the node capacitance etc. \cite{combn1_glitch_ht}.

\subsection{\label{sub:Circuit-simulation}Sample circuit simulation}

The post-layout SPICE netlist of the test circuit is the template
from which sample simulation decks are generated. To generate a sample
simulation deck, we pick a random drain \textit{(d)} of a transistor
in the design at which the glitch will be injected. The probability
of a drain being affected by a particle strike is assumed to be proportional
to its area. We then select a random clock cycle\textit{ (t)} from
the RTL reference base, and introduce the glitch at a random time
instant \textit{(k)} in the selected clock cycle.

Thus, a sample simulation deck is generated using three random numbers,
which are generated using uniform sampling from a single seed, so
that the results are reproducible.\textcolor{black}{{} This process
is followed till several simulation decks are generated with different
}\textit{\textcolor{black}{(d,t,k)}}\textcolor{black}{{} values and
are simulated using the Ngspice circuit simulator \cite{ngspice}.}

\begin{figure}[H]
\includegraphics[scale=0.45]{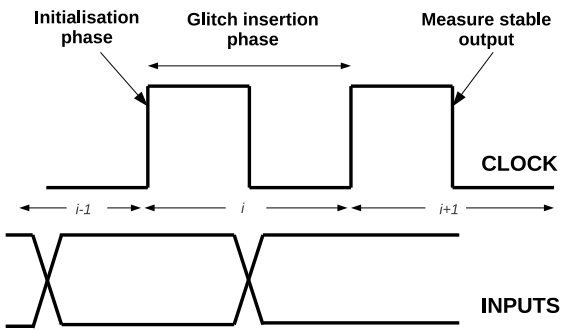}

\caption{\label{fig:Simulation-cycle}Simulation phases}
\end{figure}

The SPICE simulation is run for two and a half clock cycles, of which
the first half cycle is for initialization, the next cycle is for
glitch insertion, the following half-cycle is for monitoring the stabilized
flip-flop values and the last half-cycle is just for completeness.
This is depicted in Fig. \ref{fig:Simulation-cycle}.

In the initialisation phase, the input flip-flops of the SPICE netlist
are initialized at the positive edge of the selected clock cycle.
The following clock cycle (i\textsuperscript{th}clock cycle) is the
glitch insertion phase, in which the glitch can be randomly inserted
at any point in time. Any changes in the inputs in the RTL simulation
in this clock cycle are exactly reproduced in the SPICE simulation.
In the (i+1)\textsuperscript{th}clock cycle, the outputs of the fault-injected
circuit simulation are measured and recorded. The decks are simulated
in parallel using GNU Parallel \cite{Tange2011a} on multiple cores
on a high performance computing facility and the outputs of each faulty
simulation are saved.

\subsection{Estimation of multiple bit flip probability}

The expected values from the RTL reference simulation are compared
with those obtained from the fault-injected SPICE simulation. A table
of the resulting difference between the two values is created, in
which, a '1' indicates that the fault-injection resulted in a bit
flip and a '0' indicates that the bit did not flip. The cases which
do not have any bit flip are not of interest to us. We calculate an
estimate for the conditional probability $\theta$ and the standard
error as indicated in Eq. (\ref{eq:1}) and Eq. (\ref{eq:standardError})
respectively. Our algorithm is said to have converged when the standard
error is within 10\% of the estimate. The total number of simulations
run for each circuit varies, depending on how long it takes for the
standard error to become small enough.

\section{\label{sec:Results}Results}

Our experiments are run in parallel on a high performance computing
facility, utilizing upto a maximum of 200 cores. A maximum of 8 test-circuits
with thousands of sample decks for each circuit could be simulated
in parallel. The time taken to generate results for the largest circuit
was nearly 2 days. The number of cores used for simulations is based
on the circuit size, availability of cores and the time taken for
each simulation. The glitch injection experiments are performed on
the\textit{ ISCAS'85} benchmark circuits and a few other example circuits.
Flip-flops are added to all the inputs and outputs of the combinational
logic circuits. The clock frequency for each circuit is set to the
maximum operable frequency of the post-layout netlist, which is determined
by post-layout timing analysis. The clock frequency ranges from 90MHz
to 125MHz across all the circuits, except for the 3:8 decoder, which
was simulated at 1GHz.

\begin{table*}[t]
\caption{\label{tab:Conditional-probability-of}Conditional probability of
multiple bit flips given atleast one flip}

\begin{tabular}{c|>{\centering}p{1.7cm}|>{\centering}p{1.7cm}|>{\centering}p{1.5cm}|>{\centering}p{1.5cm}|>{\centering}p{1.6cm}|>{\centering}p{2cm}|>{\centering}p{1.6cm}}
\hline 
{\footnotesize Example} & {\footnotesize No. of gates in the design} & {\footnotesize No. of gates post pnr} & {\footnotesize No. of inputs} & {\footnotesize No. of outputs} & {\footnotesize Logic depth} & {\footnotesize No. of simulations run} & $\theta$\tabularnewline
\hline 
\hline 
{\footnotesize c432 } & {\footnotesize 160} & {\footnotesize 158} & {\footnotesize 36} & {\footnotesize 7} & {\footnotesize 3 \textendash{} 20} & {\footnotesize 63000} & {\footnotesize 12.8\%}\tabularnewline
\hline 
{\footnotesize c499 } & {\footnotesize 202} & {\footnotesize 217} & {\footnotesize 41} & {\footnotesize 32} & {\footnotesize 4 \textendash{} 16} & {\footnotesize 20000} & {\footnotesize 3.9\%}\tabularnewline
\hline 
{\footnotesize c880 } & {\footnotesize 383} & {\footnotesize 253} & {\footnotesize 60} & {\footnotesize 26} & {\footnotesize 4 \textendash{} 19} & {\footnotesize 65000} & {\footnotesize 1.27\%}\tabularnewline
\hline 
{\footnotesize c1355 } & {\footnotesize 546} & {\footnotesize 314} & {\footnotesize 41} & {\footnotesize 32} & {\footnotesize 4 \textendash{} 21} & {\footnotesize 61000} & {\footnotesize 0.01\%}\tabularnewline
\hline 
{\footnotesize c1908 } & {\footnotesize 880} & {\footnotesize 267} & {\footnotesize 33} & {\footnotesize 25} & {\footnotesize 2 \textendash{} 23} & {\footnotesize 48000} & {\footnotesize 5.20\%}\tabularnewline
\hline 
{\footnotesize c2670 } & {\footnotesize 1193} & {\footnotesize 672} & {\footnotesize 233} & {\footnotesize 140} & {\footnotesize 4 \textendash{} 28} & {\footnotesize 48000} & {\footnotesize 5.99\%}\tabularnewline
\hline 
{\footnotesize c3540 } & {\footnotesize 1669} & {\footnotesize 633} & {\footnotesize 50} & {\footnotesize 22} & {\footnotesize 5 \textendash{} 30} & {\footnotesize 40000} & {\footnotesize 15.6\%}\tabularnewline
\hline 
{\footnotesize c5315 } & {\footnotesize 2307} & {\footnotesize 978} & {\footnotesize 178} & {\footnotesize 123} & {\footnotesize 2 \textendash{} 19} & {\footnotesize 33000} & {\footnotesize 7.01\%}\tabularnewline
\hline 
{\footnotesize c7552 } & {\footnotesize 3512} & {\footnotesize 1241} & {\footnotesize 207} & {\footnotesize 108} & {\footnotesize 3 \textendash{} 27} & {\footnotesize 38000} & {\footnotesize 8.83\%}\tabularnewline
\hline 
{\footnotesize 3:8 decoder} & {\footnotesize 28} & {\footnotesize 26} & {\footnotesize 3} & {\footnotesize 8} & {\footnotesize 5} & {\footnotesize 110000} & {\footnotesize 0.70\%}\tabularnewline
\hline 
{\footnotesize 8-bit multiplier} & {\footnotesize 200} & {\footnotesize 169} & {\footnotesize 16} & {\footnotesize 16} & {\footnotesize 12 \textendash{} 15} & {\footnotesize 36000} & {\footnotesize 31.4\%}\tabularnewline
\hline 
\end{tabular}
\end{table*}

In Table \ref{tab:Conditional-probability-of}, we report the estimates
for $\theta$ for all the test-circuits. The reported value is the
minimum of the values obtained in the 95\% confidence interval. We
observe that $\theta$ ranges from 0.01\% up to as high as 31\%. Also,
the value of $\theta$ is quite significant in a majority of the circuits.
For a particular circuit, the value of $\theta$ seems to depend on
various factors such as the structure of the circuit itself, the presence
of balanced paths, logic depth, the kind of logic gates used and the
input combinations. This needs further study. However, it is clear
that for most circuits, a single bit-flip model is not an adequate
fault model for soft errors, if realistic estimates of reliability
are to be obtained. Further, these observations are based on detailed
post-layout circuit simulations without any simplifying assumptions
(other than the glitch injection itself, which is assumed to be uniform
across the circuit, and is assumed to affect a single drain). Thus,
we have a high degree of confidence in the conclusions.

Although we report results for designs at 180nm technology, we expect
that $\theta$ is likely to increase with technology scaling as indicated
in \cite{mult_combn_scaling}. This is due to the possibility of a
single strike affecting multiple nearby gates resulting in multiple
transients. Hence, a multiple bit-flip fault model for soft errors
is likely to be necessary at lower technology nodes as well.

\section{\label{sec:Conclusions}Conclusions}

In this paper, we have demonstrated that a single bit-flip model is
not adequate to model soft-errors in a logic circuit. We use Monte-Carlo
sampling with detailed post-layout circuit simulations to estimate
the conditional probability that multiple bits flip given that a single
bit has flipped. This conditional probability is estimated for a variety
of test-circuits, and significant values upto as high as $0.31$ (for
a multiplier) are observed. A wide variation is seen across the test-circuits,
indicating that this probability depends on the structure and logic
functionality of the individual circuits. However, the broad conclusion
is that, multiple bit-flips must necessarily be considered in order
to obtain a realistic fault model for soft errors in logic circuits.

\bibliographystyle{plain}

\end{document}